\documentclass[conference]{IEEEtran}
\IEEEoverridecommandlockouts

\def\IEEEtitletopspace{18pt}
\usepackage{cite}
\usepackage{amsmath,amssymb,amsfonts}
\usepackage[margin=0.75in]{geometry}
\usepackage{graphicx}
\usepackage{array}
\usepackage{booktabs}
\usepackage{longtable}
\usepackage{subcaption}
\usepackage{hyperref}
\usepackage{textcomp}
\usepackage{xcolor}
\usepackage{algorithm}
\usepackage{algpseudocode}
\usepackage{booktabs}
\usepackage{multirow}

\def\BibTeX{{\rm B\kern-.05em{\sc i\kern-.025em b}\kern-.08em
    T\kern-.1667em\lower.7ex\hbox{E}\kern-.125emX}}
\begin{document}

\title{Automated Route-based Conflation Between Linear Referencing System Maps And OpenStreetMap Using Open-source Tools\\
}

\author{ Gibran~Ali,
Neal~Feierabend,
Prarthana~Doshi\\
Whoibin~Chung,
Simona~Babiceanu,
Michael~Fontaine
\thanks{G. Ali*, N. Feierabend, P. Doshi are with  Virginia Tech Transportation Institute (*email: gali@vtti.vt.edu)}
\thanks{W. Chung, S. Babiceanu, and M. Fontaine are with the Virginia Department of Transportation}
}

\maketitle

\begin{abstract}
Transportation researchers and planners utilize a wide range of roadway metrics that are usually associated with different basemaps. Conflation is an important process for transferring these metrics onto a single basemap. However, conflation is often an expensive and time-consuming process based on proprietary algorithms that require manual verification. 

In this paper, an automated  open-source process is used to conflate two basemaps: the linear reference system (LRS) basemap produced by the Virginia Department of Transportation and the OpenStreetMap (OSM) basemap for Virginia. This process loads one LRS route at a time, determines the correct direction of travel, interpolates to fill gaps larger than 12 meters, and then uses Valhalla’s map-matching algorithm to find the corresponding points along OSM’s segments. Valhalla’s map-matching process uses a Hidden Markov Model (HMM) and Viterbi search-based approach to find the most likely OSM segments matching the LRS route.

This work has three key contributions. First, it conflates the Virginia roadway network LRS map with OSM using an automated conflation method based on HMM and Viterbi search. 
Second, it demonstrates a novel open-source processing pipeline that could be replicated without the need for proprietary licenses. 
Finally, the overall conflation process yields over 98\% successful matches, which is an improvement over most automated processes currently available for this type of conflation. 
\end{abstract}


\section{Introduction}
 
There are several basemaps commonly used by traffic planners and researchers. For example, United States (US) state departments of transportation (DOTs) and the US Federal Highway Administration both use linear referencing system (LRS) basemaps to catalog the roadway infrastructure \cite{hausman2014all}. These maps include several types of attributes for each segment, such as roadway geometry, functional class, speed limits, and annual average daily traffic. Similarly, mapping services---such as OpenStreetMap (OSM), TomTom, and HERE---all maintain separate basemaps with various attributes, such as road class,  travel time, reference speeds, curvature, and grade. 

Utilizing attributes from different basemaps is a key challenge for Intelligent Traffic System (ITS) performance evaluation and Traffic safety analysis. Basemaps can differ in many ways. First, the geometry describing the roadway segments is often not aligned in different maps due to differences in the source of the information. This can be both due to the roadway centerline being different as well as the segments being broken up in different places. Second, different maps may use different underlying assumptions about  roadway segments, with some splitting bi-directional roads into different geometries and others combining them onto a single centerline. Finally, different maps use different logic to break roadway segments. Some may break segments only at major changes in properties, while others keep roadway segments below specific length thresholds even when roadway properties do not change. This often means that the coordinates used to describe the geometry can be far apart even when they are along the same line. 

Therefore, conflation of different basemaps is an important step that ITS research and planning groups often perform to combine roadway metrics. However, conflation can be computationally intensive, error prone, and difficult to verify. Additionally, new basemaps are released periodically, resulting in the need for ongoing conflation efforts. Therefore, many ITS groups spend considerable resources conflating multiple basemaps at regular intervals.

Several methods have been proposed to automate conflation. However, to the best of the authors' knowledge, there are no published studies in the existing literature that conflate an LRS basemap with OSM using open-source tools. 

In this paper, an automated method is described that conflates the Virginia DOT's (VDOT's) LRS basemap with OSM's basemap. The conflation results show that over $98\%$ of roadway segments were satisfactorily matched.  This pipeline is based on open-source tools that any ITS group could apply for their own conflation applications. Several strategies are discussed in the pre-processing and cleaning steps to improve conflation results. Additionally, the original datasets, along with the conflation results, are made available for review and an interactive tool is created to visualize the conflation results.

\section{Background}
\subsection{Need for Map Conflation}

In the context of roadway networks, the LRS is a set of procedures for retaining information of specific points or sections along a route using a linear measure, such as a milepost \cite{hausman2014all}. Over the past decade, enhancements in GIS software and federal guidance have increased the adoption of LRS-based mapping by state DOTs. In 2012, the US federal government required all US state DOTs to develop and submit an LRS covering all public roads as part of the Highway Performance Monitoring System (HPMS) submittal---which is known as the "All Roads Network of Linear Referenced Data (ARNOLD)" requirement \cite{hausman2014all}. A key advantage of the LRS model is dynamic segmentation, which decouples the spatial geometry from attributes data. This capability ensures the storage and retrieval of various types of data such as signs, maintenance records, crash sites, and roadway parameters  without having to break and re-digitize the geometry for every thematic layer. Consequently, most state DOTs and transportation planning agencies now use some form of LRS-based roadway maps for data analysis and reporting, although the criteria and methods used to create these maps may differ from state to state. 

ITS researchers and practitioners often require additional information such as travel time estimates and roadway attributes like curvature and speed limits, which are usually maintained in separate roadway basemaps. For example, planners may use data from HERE, TomTom, and OSM for these types of analysis. However, each of these map providers maintains its own distinct basemap that would need to be conflated with the LRS map used by ITS agencies.

Over the past decade, large-scale vehicle trajectory datasets have been increasingly used by ITS planners to understand driver behavior on roadways. These datasets typically contain detailed information about vehicle location, kinematics, and other states recorded at regular intervals. However, these trajectories need to be map-matched to a basemap for the data to be useful for ITS applications. While many proprietary map providers offer their own map-matching services, the costs associated with processing large-scale datasets can be prohibitively high. A free-of-cost solution is to map-match such datasets using open-source basemaps and tools. OSM is a crowd-sourced, freely available basemap. Similarly, Valhalla is an open-source software that can be used freely by setting up a local instance. 

\subsection{Conflation Methodologies}

Map conflation is a critical process in geospatial data integration, where multiple spatial datasets are merged to transfer attributes across maps \cite{saalfeld1988conflation}. Over the years, various methodologies have emerged to address the complexities of map conflations. Notable examples include NetMatcher  by Mustière et al. \cite{mustiere2008matching}, Delimited Stroke Oriented (DSO) by Zhang et al.\cite{zhang2008delimited}, and SimMatching by Schafers et al. \cite{schafers2014simmatching}, all of which use greedy heuristic approaches that stand out for quick results using local optimization strategies. Similarly, Volker et al. \cite{walter1999matching}, Li et al. \cite{li2011optimisation}, and Tong et al. \cite{tong2014linear} are examples of optimization-based conflation strategies that aim to find the best possible solution using an objective function. While these methods have shown considerable progress in the field, they tend to be computationally expensive, especially for large-scale datasets. This underscores the need for new and improved approaches.

\begin{figure}[t]
    \centering
    \includegraphics[width=0.95\columnwidth,keepaspectratio]{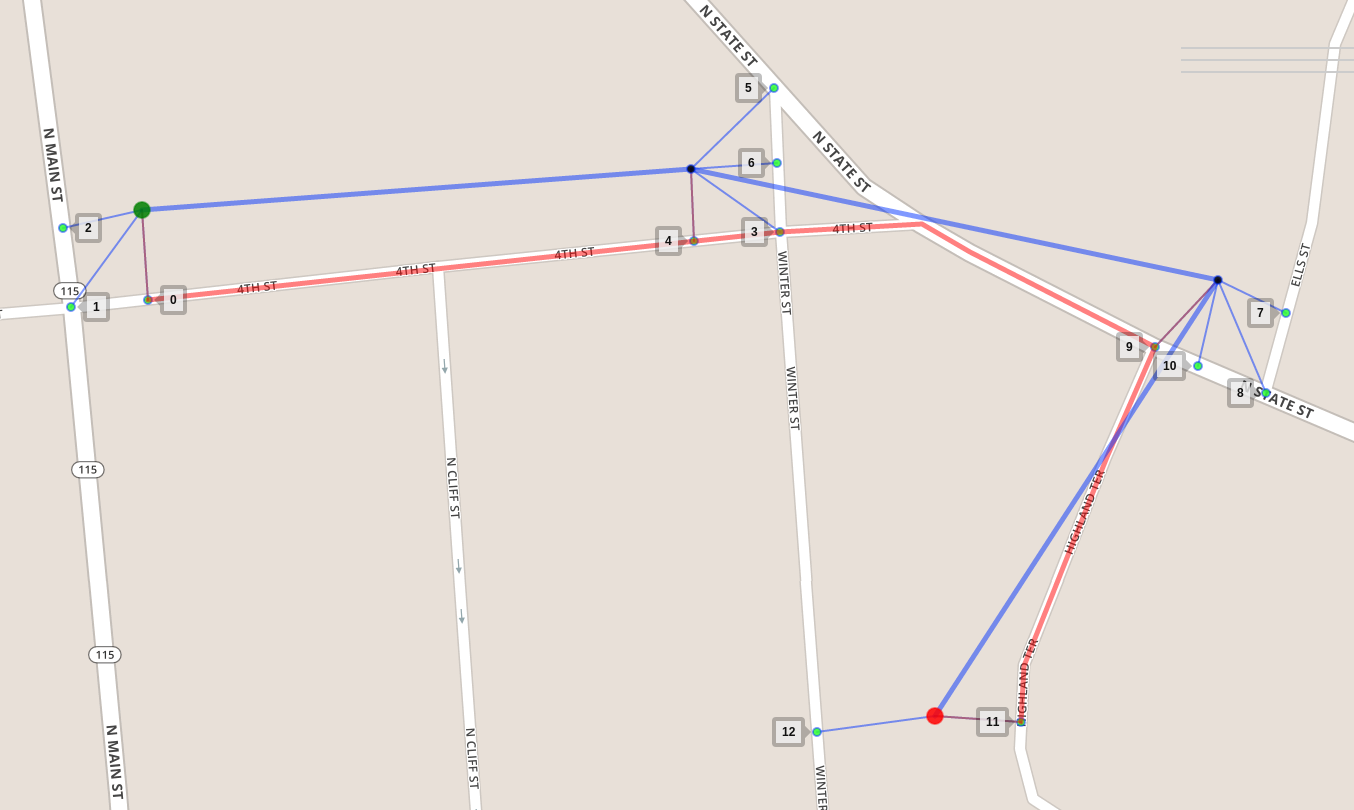}
\caption{Valhalla algorithm approach using HMM to conflate noisy GPS data to roadway network. The blue line starting from green point to the red point represents the raw data and the red line represents the matched roadway. \cite{valhalla_contributors_valhalla_2025} (Licensed under MIT License from \href{https://github.com/valhalla/valhalla/blob/master/docs/docs/meili/algorithms.md}{Valhalla GitHub repository}).}
\label{fig:vahla-algo-gps}
\end{figure}

Newson et al.\cite{newson2009hidden} introduced a significant advancement in map-matching and conflation algorithms using Hidden Markov Models (HMMs) to identify the most probable sequence of road segments corresponding to a given series of GPS points. The Valhalla libraries used in this paper also rely on this approach. Figure \ref{fig:vahla-algo-gps} shows how Valhalla conflates noisy GPS data---shown as the green to black to red dots---and aligns them to the most likely road network path. For a given sequence of points, the algorithm selects potential match candidates represented as points 0 to 12 and then chooses the best fit---candidates 0, 4, 9, and 11---by finding the optimal path using HMM and Viterbi algorithm approaches \cite{valhalla_contributors_valhalla_2025}.

Valhalla is an open-source routing engine that provides multiple services, such as map-matching, path-finding, and maneuver-based navigation, all built on top of the OSM's basemap \cite{saki2022practical}. The Valhalla map-matching algorithm, Meili, is primarily designed for aligning noisy vehicle trajectory data consisting of GPS points recorded at regular intervals—to the underlying road network. In this paper, we demonstrate that this algorithm is also well suited to conflating an LRS map with a state DOT's map of the same region. 

\subsection{Datasets Used}
This paper conflates two basemaps: the VDOT LRS map and OSM's basemap representing roadways in Virginia. There are multiple versions of the VDOT LRS basemap, and for this project, the edge-route overlap LRS was selected. This map consists of 1.78 million  features ($\approx$620,000 unique) representing roadway segments along various routes.  The ``overlap'' in the name implies that a single roadway segment could be represented multiple times due to overlapping, collinear routes.

Some key pieces of information associated with each roadway feature in the VDOT LRS are:
\begin{enumerate}
    \item \textbf{Edge Route Key}: A unique identifier for each edge. 
    \item \textbf{Route Name}: A unique identifier for each route. The route name indicates whether the route is prime or non-prime as well as the cardinal direction of the route.
    \item \textbf{Master Route Name}: The master route is the definitive route on any edge upon which all event data is recorded. There can only be one master route for a given edge.
    \item \textbf{Route Category}: This metric provides information about the roadway class or category.
    \item \textbf{Route Transport Edge Sequence}: The edge sequence number provides information about the order of edges in a route. All the edges belonging to a route can be ordered correctly using this metric.
    \item \textbf{Geometry}: Each edge consists of several points with X, Y, and M values. The X and Y values represent the longitude and latitude respectively of each point, while the M value represents the measured distance along the route in miles from the begining of the route.
\end{enumerate}

 There are 502,099 routes and 194,439 master routes in the VDOT LRS. The master routes themselves represent 619,284 features with a total length of 103,463 miles.
The OSM roadway dataset for Virginia consists of 338,421 features, where each feature uniquely represents a roadway segment without overlaps. The actual number of features can be much higher (\textgreater 1.5 million), but for the purpose of this analysis, we chose only those segments that were most likely to be public paved roadways. Each roadway feature consisted of a unique identifier called osm\_id, a highway tag that classifies the roadway type, a geometry or shape consisting of latitude and longitude points, and a tag column that can store information about the roadway segment in a key-value format. The OSM dataset covers 107,930 miles, with an average feature length of 0.31 miles, compared to 0.17 miles for the VDOT LRS. Therefore, on average, the OSM segments are much longer than the VDOT LRS segments.

\begin{figure}[t]
    \centering
     \begin{subfigure}[b]{\columnwidth}
        \centering
        \includegraphics[width=\columnwidth,keepaspectratio]{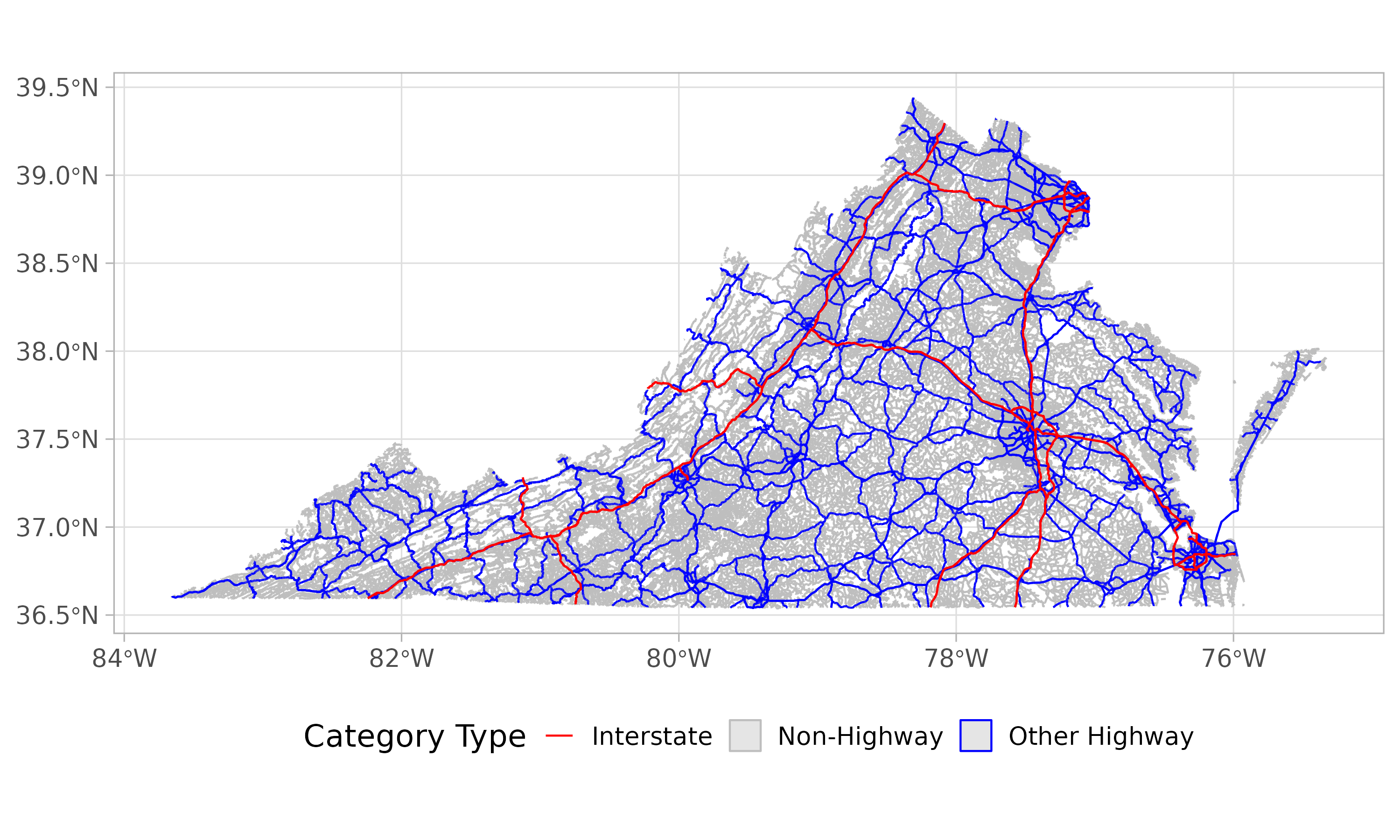}
        \caption{VDOT LRS.}
    \end{subfigure}
     \hfill
      \begin{subfigure}[b]{\columnwidth}
        \centering
        \includegraphics[width=\columnwidth,keepaspectratio]{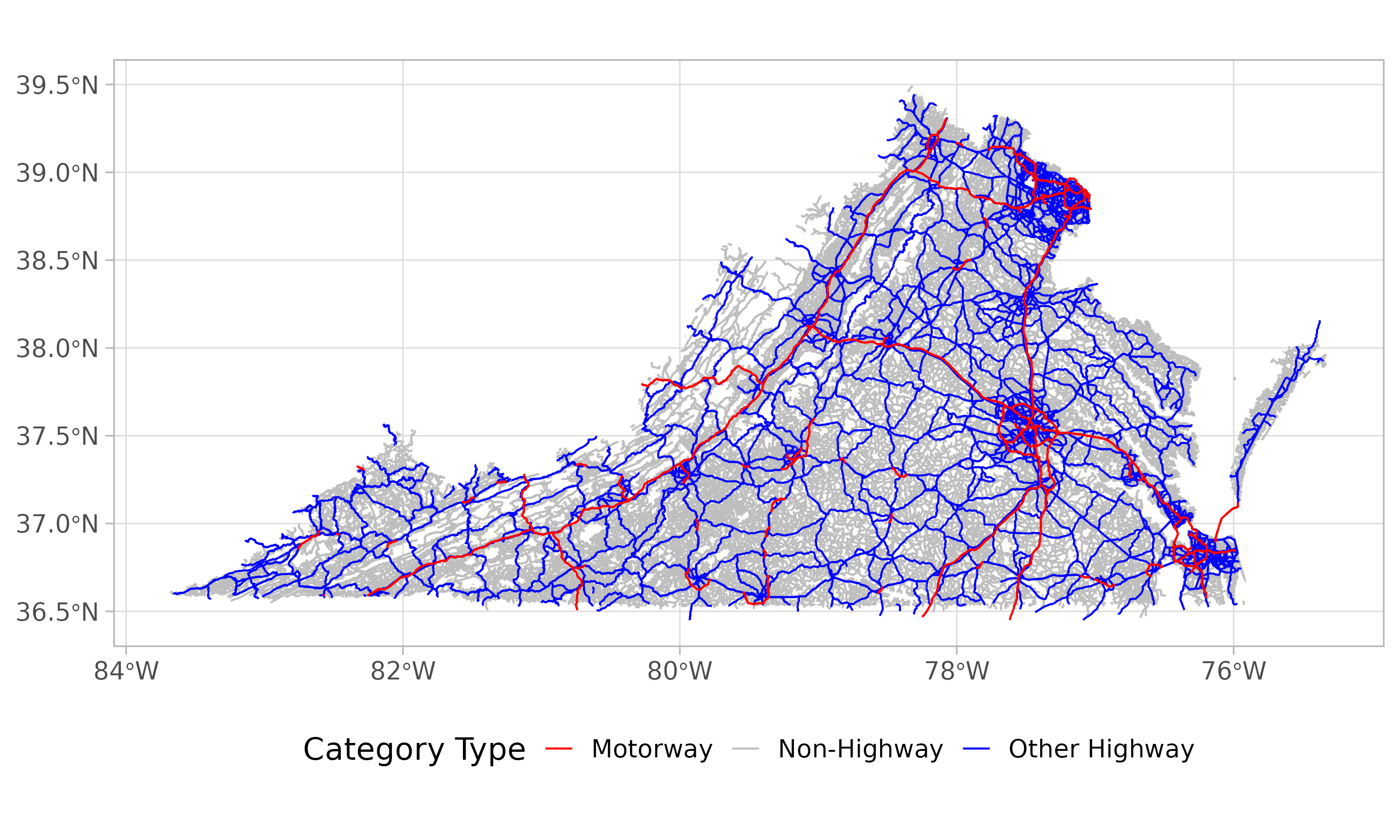}
        \caption{OpenStreetMap.}
    \end{subfigure}
\caption{Comparison of VDOT LRS and OSM basemaps for Virginia.}
\label{fig:roadway-maps}
\end{figure}

Figure \ref{fig:roadway-maps} compares the LRS and OSM maps across Virginia. The Category Type shown in the figure is derived from the route category variable in the VDOT LRS and from the highway tag in the OSM dataset. The two sources do not have one-to-one mapping, and the visualization is purely for illustration purposes. The figure clearly shows that, overall, the roadway network represented by the two maps is nearly identical, with differences arising from categorical classifications.

\section{Methodology}

Algorithm \ref{table:algorithm} outlines the step-by-step conflation process applied to a single route. The same process was followed for each of the 501,827 routes in the VDOT LRS using a parallel processing approach. The overall methodology can be broken down into three parts: pre-processing, map-matching using Valhalla, and post-processing cleanup and packaging. 

\begin{algorithm}[h]
\caption{Route-Based Conflation Process}
\label{table:algorithm}
\begin{algorithmic}
\Procedure{ProcessRoute}{route}
    \State Load a single route from the edge-based route overlap dataset
    \ForAll{edges in route}
        \State Decompose LRS geometry into \texttt{(X, Y, M)} values
    \EndFor
    \State Calculate distance between each pair of consecutive points
    \If{consecutive point distance  $>$ 12 meters}
        \State Interpolate intermediate points at 10-meter intervals
    \EndIf
    \State Determine route's cardinal direction and prime status using route name
    \If{route is non-prime \textbf{or} southbound \textbf{or} westbound}
        \State Multiply all \texttt{M} values by $-1$
    \EndIf
    \State Sort the entire route by ascending \texttt{M} values
    \If{Number of points in routes $\geq$ 70,000}
        \State Create batches of 50,000 points and move to next step
    \EndIf
    \State Prepare request body and submit to Valhalla API through \texttt{trace\_attributes} with \texttt{map\_snap} option
    \If{response is successful}
        \State Save response data
    \Else
        \State Log failure 
    \EndIf
    \State Clean route data for segments with fold back on themselves
    \If{route is non-prime \textbf{or} southbound \textbf{or} westbound}
        \State Multiply min and max \texttt{M} values by $-1$
        \State Swap min and max positions
    \EndIf
    \ForAll{unique \texttt{(edge\_rte\_key, osm\_id)} combinations}
        \State Find minimum and maximum \texttt{M} values
    \EndFor
\EndProcedure
\end{algorithmic}
\end{algorithm}

The pre-processing step transforms each VDOT LRS route into a format suitable for conflation with OSM. For each route, the LRS feature is converted from a linestring to the constituent latitude, longitude, and milepost values. If two consecutive points within a route are further than 12 meters away from each other, the distance is interpolated at an interval of 10 meters. Another important step during the pre-processing stage is to correctly orient the roadway geometry coordinates in the direction of travel. Since the Valhalla map-matching algorithm was originally developed for matching noisy GPS data from vehicle trajectories to OSM roadways, it is sensitive to travel direction, and it is therefore important that the input to the algorithm is in the correct order. 

For routes in the VDOT LRS that are non-prime (southbound and westbound), the order of points is opposite to the direction of travel. Therefore, the milepost value of these points is multiplied by -1 to reverse their order. This is specific to VDOT's LRS and may be different for other LRS maps. 

Once the points are correctly interpolated and ordered, they are submitted to the Valhalla API through the \texttt{trace\_attributes} with \texttt{map\_snap} option. The \texttt{map\_snap} option specifies that the HMM-based map-matching method discussed earlier is applied. The authors set up a local deployment of the Valhalla API on five servers with  2-CPU thread and 16 GB of memory each. The submission of routes was set up on a parallel compute cluster consisting of 48 nodes with 24 CPU cores and 256 GB of memory each. Over 50 parallel jobs were set up and the load was balanced so that the number of parallel jobs did not exceed the capacity of Valhalla instances. The Valhalla instance settings can be modified to adjust for the maximum number of points that can be matched at a time. For this study, the maximum number of points was set around 100,000, but every time the route exceeded 70,000 points, it was broken up and processed in batches of 50,000 points.

The output from the Valhalla API provides a matched coordinate along the centerline of an OSM roadway for every submitted coordinate. The output also includes a corresponding unique roadway identifier, called OSM\_id, for each coordinate. This output is then reprocessed to create a table with the original route name, edge route key, submitted coordinate, interpolated milepost value, returned coordinate, OSM\_id, and distance between the set of coordinates. This table then goes through post-processing steps that remove improper matches.

Finally, for each combination of route name, edge route key, and OSM\_id, the minimum and maximum values of the milepost are summarized. The milepost values are again multiplied by -1 and swapped for routes that are ordered in the opposite direction of travel. This final dataset is a conflation key that provides  mapping between the VDOT LRS and the OSM basemap.

\section{Results}

Table \ref{tab:results-summary} summarizes the results of the conflation process. Out of 1.78 million edges representing 502,029 routes, 99.83\% were successfully processed through the pipeline. Only 0.17\% of the routes did not match; upon examination, most of these were due to the OSM not having corresponding roadways for these routes. When the results for unique roadway segments without overlap were examined by selecting the master route edges only, similar statistics were seen: 99.79\% of the edges and  99.87\% of the mileage were processed successfully. However, these initial statistics only indicate the success of processing, not the quality of the matches.

\begin{table}[]
\centering
\caption{Summary of the conflation results.}
\label{tab:results-summary}
\resizebox{\columnwidth}{!}{%
\begin{tabular}{@{}lll@{}}
\toprule
Type                                     & Metric                        & Value               \\ \midrule
\multirow{8}{*}{All edges}               & Total edges matched           & 1,780,490 (99.83\%) \\
                                         & Total mileage matched         & 321,685 (99.91\%)   \\
                                         & Edges $<$ 12 $\bar{x}$         & 1,752,341 (98.28\%)   \\
                                         & 50th \%ile $\bar{x}$ & 2.5 m               \\
                                         & 75th \%ile $\bar{x}$ & 2.9 m               \\
                                         & 90th \%ile $\bar{x}$ & 3.6 m               \\
                                         & 95th \%ile $\bar{x}$ & 4.9 m               \\
                                         & 98th \%ile $\bar{x}$ & 9.9 m               \\
                                         & 99th \%ile $\bar{x}$ & 18.3 m              \\\midrule
\multirow{8}{*}{Master route edges only} & Total edges matched           & 617,988 (99.79\%)   \\
                                         & Total mileage matched         & 103,332 (99.87\%)   \\
                                         & Edges $<$ 12 $\bar{x}$         & 605,622 (97.8\%)   \\
                                         & 50th \%ile $\bar{x}$ & 2.5 m               \\
                                         & 75th \%ile $\bar{x}$ & 3.0 m               \\
                                         & 90th \%ile $\bar{x}$ & 3.9 m               \\
                                         & 95th \%ile $\bar{x}$ & 5.5 m               \\
                                         & 98th \%ile $\bar{x}$ & 12.4 m             \\
                                         & 99th \%ile $\bar{x}$ & 20.6 m             \\ \bottomrule
\end{tabular}%
}
\end{table}

\begin{figure}[t]
    \centering
    \includegraphics[width=01\columnwidth,keepaspectratio]{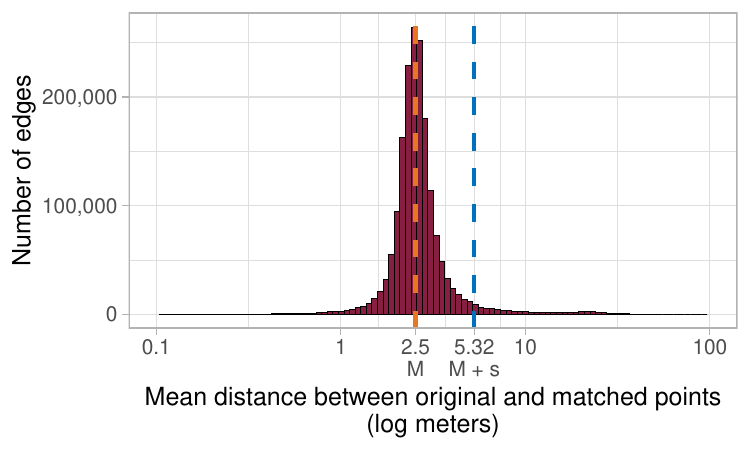}
\caption{Distribution of mean distance between original and matched GPS points. The orange vertical line represents the median and the blue vertical line represents the median $+$ standard deviation measures.}
\label{fig:distribution-mean-overall}
\end{figure}

To ascertain the quality of the matches, three approaches were used. First, the mean distance, $\bar{x}$, between the original and the matched points was calculated for each edge. Figure \ref{fig:distribution-mean-overall} shows the distribution of $\bar{x}$ and Table \ref{tab:results-summary} lists key percentile metrics of the distribution. In both types of metrics, the median $\bar{x}$ is around 2.5 meters and 98\% of the edges have an $\bar{x}$ of $\approx$ 12 meters.

\begin{figure*}[!p]
    \centering
     \begin{subfigure}[b]{0.9\textwidth}
        \centering
        \includegraphics[width=0.9\textwidth,keepaspectratio]{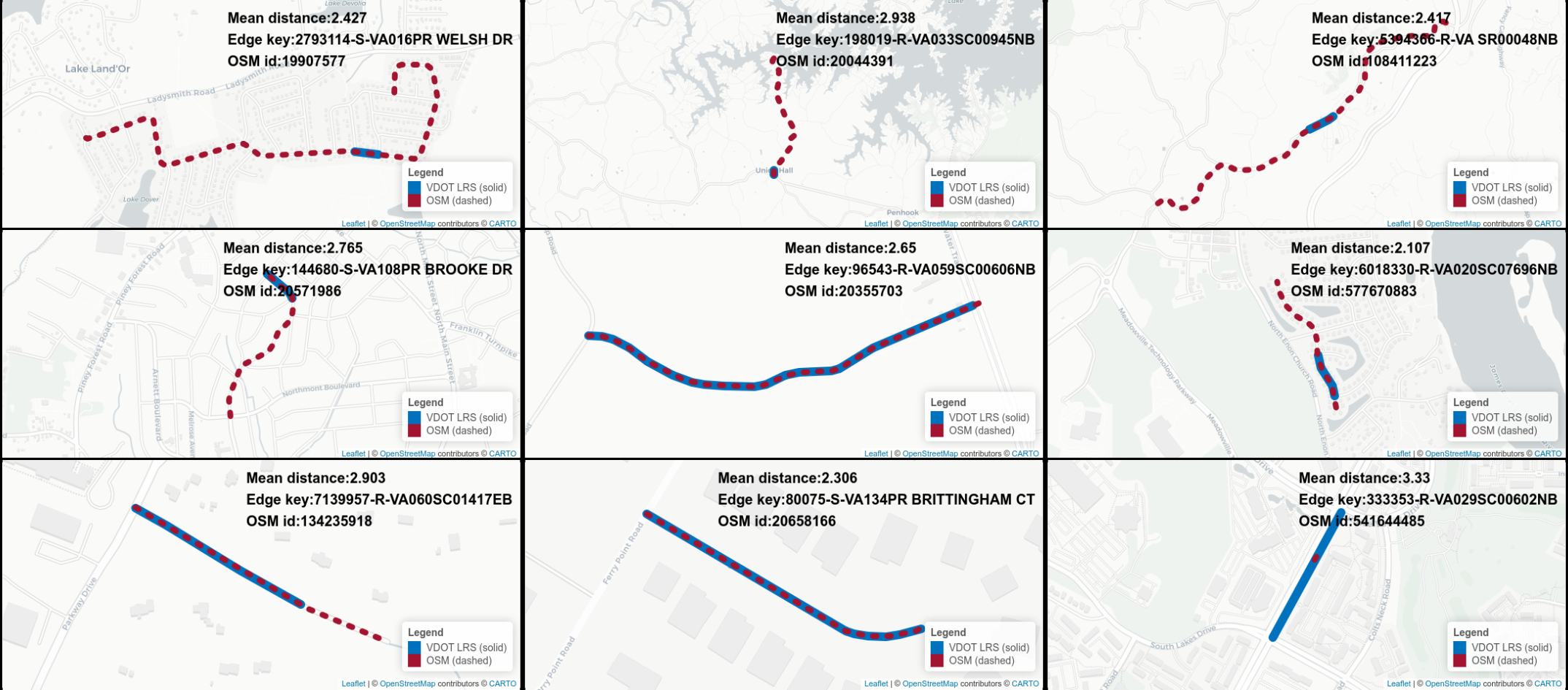}
        \caption{$0 \leq \bar{x} \leq 6$.}
    \end{subfigure}
     \hfill
      \begin{subfigure}[b]{0.9\textwidth}
        \centering
        \includegraphics[width=0.9\textwidth,keepaspectratio]{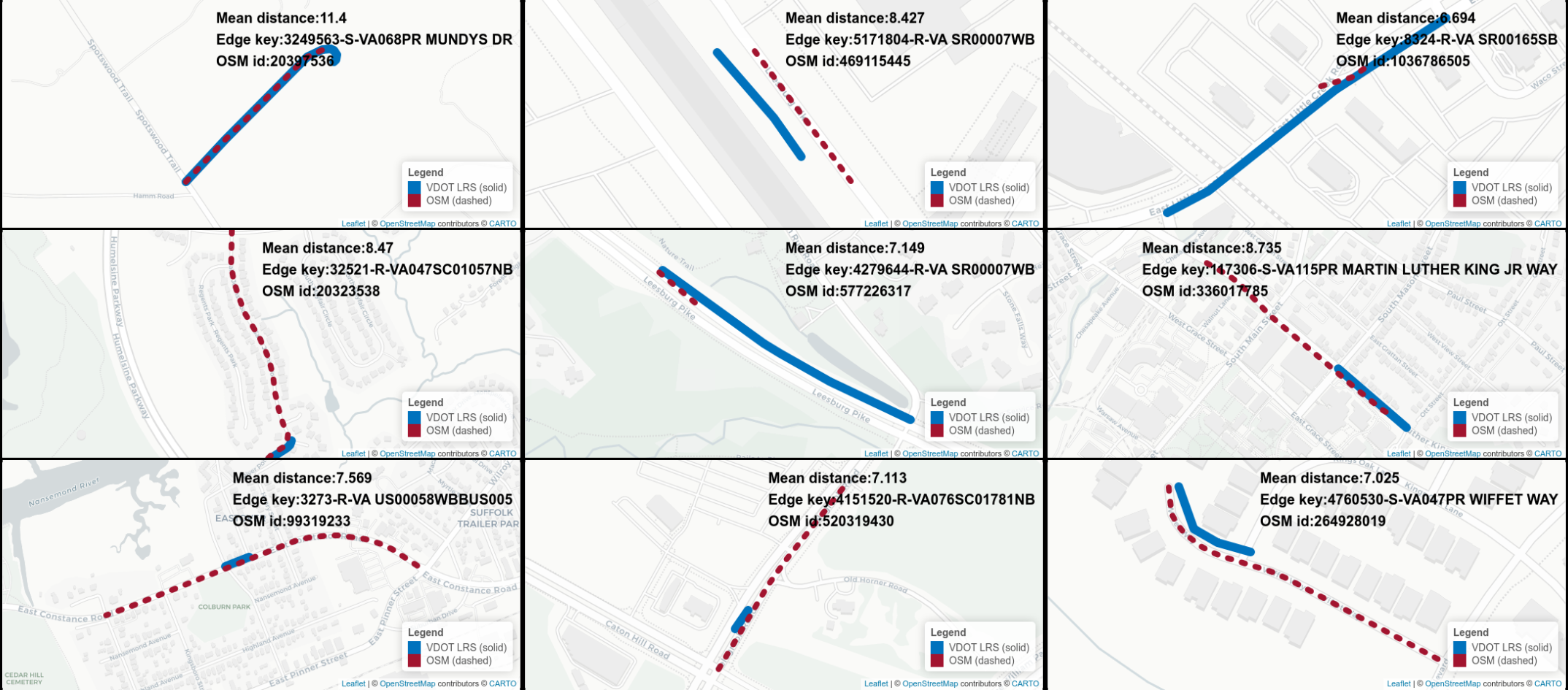}
        \caption{$6 \leq \bar{x} < 12$.}
    \end{subfigure}
     \hfill
      \begin{subfigure}[b]{0.9\textwidth}
        \centering
        \includegraphics[width=0.9\textwidth,keepaspectratio]{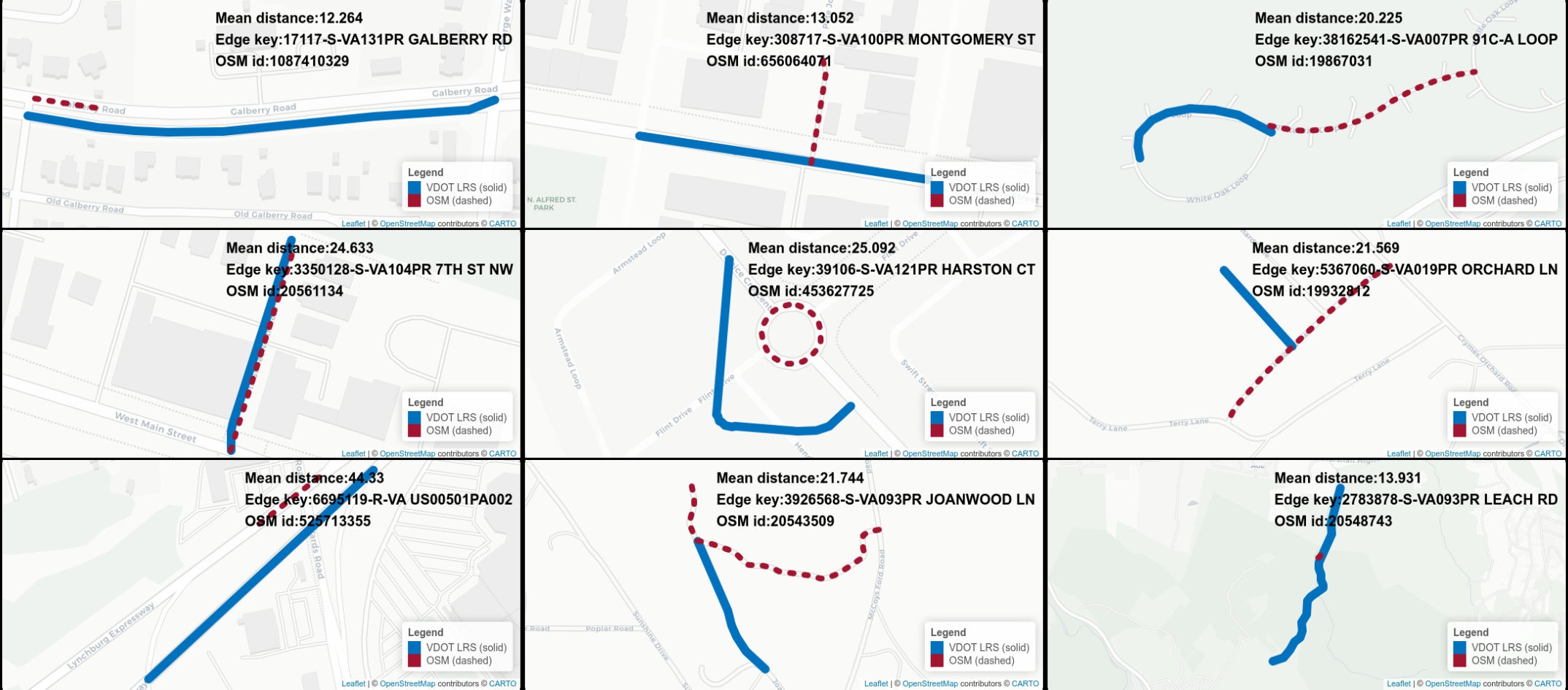}
        \caption{$12 \geq \bar{x} $.}
    \end{subfigure}
\caption{Randomly sampled edges visualizing the VDOT LRS (Blue) and OSM (Red) matching roadway segments.}
\label{fig:random-viz-match}
\end{figure*}

The second approach to ascertain the quality of metrics was to randomly sample edges from different regions of the $\bar{x}$ distribution and visualize the LRS and OSM geometries. Figure \ref{fig:random-viz-match} illustrates this approach by visualizing nine randomly sampled edges from three regions of $\bar{x}$:  $0 \leq \bar{x} \leq 6$, $6 \leq \bar{x} < 12$, and $\bar{x} \geq \ 12$. Each of the images in the grid shows the VDOT LRS edge in solid blue and the matching OSM roadway segment in dashed red. This visualization illistrates the complete geometry of both the roadway segments and not just the matched sections. However, only the overlapping sections should be compared, as multiple geometries from one dataset could have been conflated with a single geometry of the other. It should also be noted that each of these maps is at a different scale zoomed to fit the complete geometries, and therefore some distances may appear greater because of map scale.

\begin{figure*}[!t]
    \centering
    \includegraphics[width=0.9\textwidth,keepaspectratio]{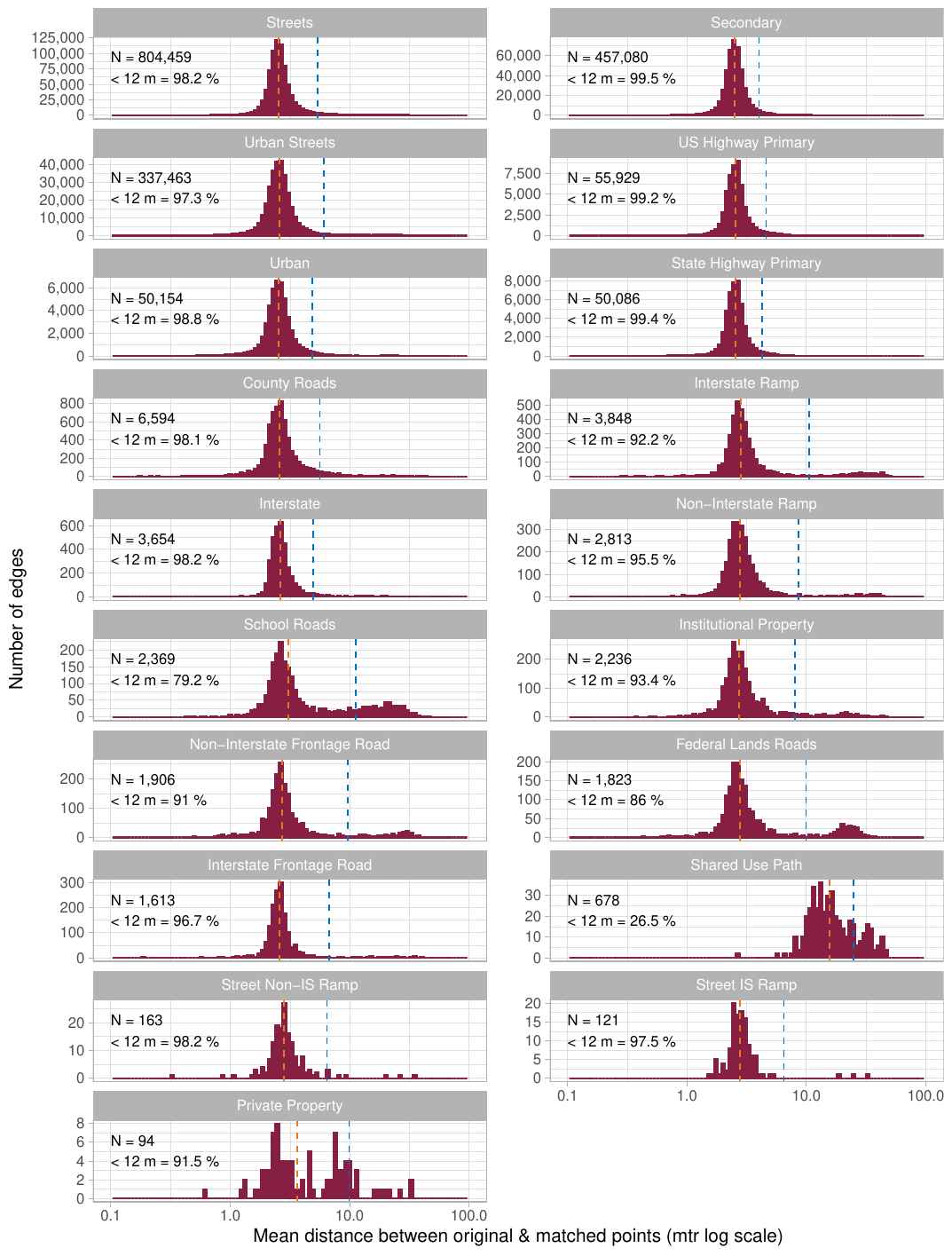}
\caption{The distribution for  mean distance between original and matched points ($\bar{x}$ grouped by the VDOT LRS route category name.}
\label{fig:distribution-mean-by-cat}
\end{figure*}

This approach shows that most matches within $0 \leq \bar{x} < 12$ are the best conflation matches for the two datasets. Some matches for $12 \geq \bar{x}$ may be acceptable, but generally need further review for refinement. Another similar approach used for determining quality of matches is illustrated in Figure \ref{fig:distribution-mean-by-cat}. This illustration shows that, for most route types with a significant number of edges, the conflation process performs well. In fact, this type of verification enabled the detection of various types of failure modes  that have since been resolved using various fixes. Currently, School Roads, Federal Lands Roads, and Shared Use Paths show failures, but since the overall number of roads under these categories are quite low, they have not been resolved.

\begin{figure*}[!t]
    \centering
    \includegraphics[width=0.85\textwidth,keepaspectratio]{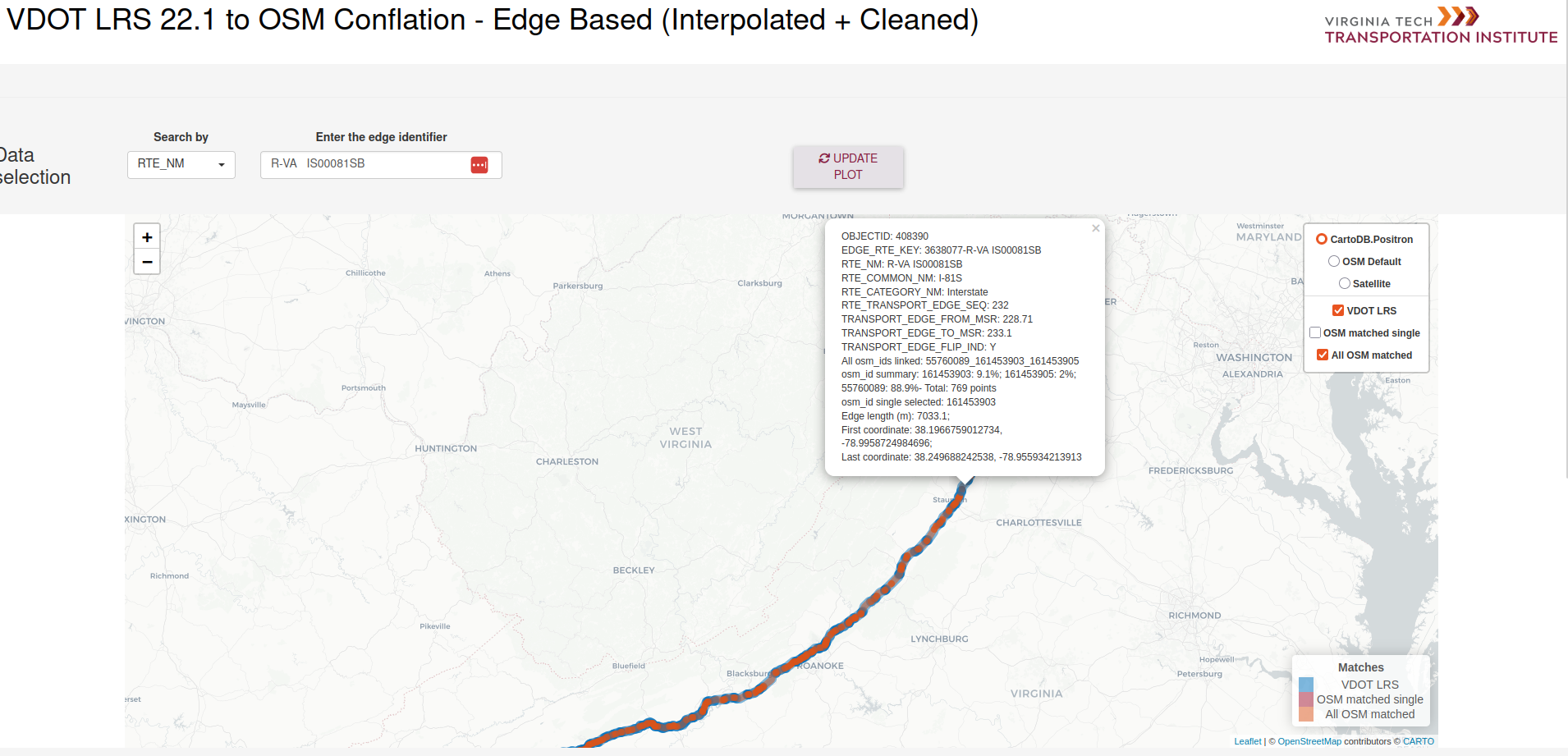}
\caption{Open interactive visualization tool to illustrate results of conflation. Accessible \href{https://dataviz.vtti.vt.edu/conflation-app-edge-based-interpolated-cleaned/}{online}.}
\label{fig:viz-tool}
\end{figure*}

Finally, the third approach for determining conflation quality  for most major routes was manual review. This approach showed that most major interstates and highways were well matched. Figure \ref{fig:viz-tool} shows an open tool developed to visualize geometry and conflation results of one complete route at a time, which can be accessed \href{https://dataviz.vtti.vt.edu/conflation-app-edge-based-interpolated-cleaned/}{online}.

The original VDOT LRS data, OSM basemap, and conflation key have been made available through a GitHub repository.

\begin{figure}[h]
    \centering
    \includegraphics[width=0.5\columnwidth,keepaspectratio]{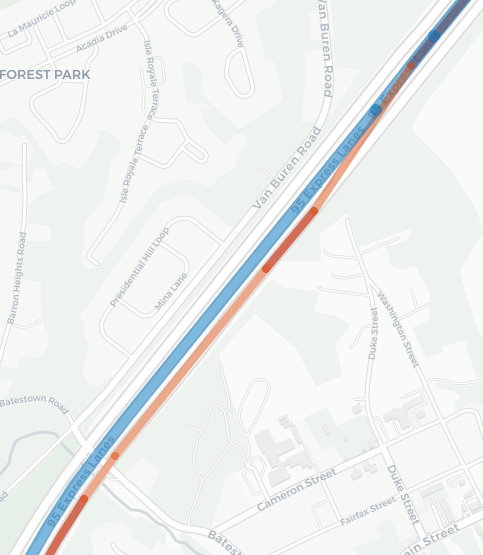}
\caption{An example of conflation failure involving reversible expressway (Route name: R-VA   IS00095RV).}
\label{fig:rev-exp}
\end{figure}

During these three different types of data quality checks, multiple limitations were also discovered. Some routes that include reversible expressways did not match perfectly even though the expressway lanes were available in the OSM dataset. Figure \ref{fig:rev-exp} shows one such example, where the VDOT LRS route R-VA   IS00095RV (in blue) does not conflate to the correct OSM roadway segments (in orange). Another limitation discovered in the review process involved the conflation of some points along a highway to service roadways. Most such issues were resolved in the post-processing cleaning steps.

\section{Discussion}

Most U.S. DOTs require conflation services at regular intervals to update their LRS maps with a wide variety of roadway parameters. These methods are often used to conflate the DOT-maintained LRS with various other open and proprietary basemaps provided by companies such as OSM, HERE, and TomTom. However, conflation is often an expensive and error-prone process. There is a need for non-proprietary conflation methods that can scale to large datasets and be repeatable without the need for manual labeling.

The work presented in this paper establishes a reliable, open-source, and scalable method to conflate the VDOT LRS map with the corresponding OSM map for Virginia. This method successfully conflated over 98\% of the 1.78 million roadway segments described in the VDOT LRS with an average distance between original and matched segments of less than 12 meters. The conflation results were checked using summary metrics, manual review of randomly selected edges, and manual review of important routes.

This work will enable the transfer of metrics collected on OSM to the VDOT LRS \cite{Ali2025ITSCTelematics}. Given that OSM is a commonly used map-matching basemap for analyzing vehicle trajectories, the output of this project will be quite beneficial to ITS research and planning activities. The conflation methods presented in this paper will also enable richer data analysis involving naturalistic driving data by providing more roadway parameters to be associated with driving data \cite{ali_characterizing_2023, gali_2021, ali_quantifying_2021, Beale2025ITSCConflation}.

\section{Future Work}

The authors have identified three major areas of future work. First, better pre-processing and cleaning steps can further improve the results of this conflation process. Second, the limitations related to reversible expressways need to be better understood and can be addressed in future work. Finally, this conflation method can be generalized to other basemaps that are often used by the transportation community.


\section*{Acknowledgment}
This work was funded by Virginia Transportation Research Council award - VTRC 121564/RC00155. The authors would also like to acknowledge various members of the Virginia Department of Transportation who provided guidance on the VDOT LRS, reviewed results, and provided feedback to improve this work.

\section*{Data and Resources}
This paper is accompanied with an interactive visualization tool (\href{https://dataviz.vtti.vt.edu/conflation-app-edge-based-interpolated-cleaned/}{https://dataviz.vtti.vt.edu/conflation-app-edge-based-interpolated-cleaned/}) and an open data repository (\href{https://github.com/gibran-ali/vdot-lrs-conflation}{https://github.com/gibran-ali/vdot-lrs-conflation}). Please allow a few minutes for the interactive data visualization to load.

\bibliography{references.bib}
\bibliographystyle{ieeetr}

\vspace{12pt}

\end{document}